\documentclass[11pt]{article}
\usepackage[a4paper,margin=1in]{geometry}
\usepackage{graphicx}
\usepackage{hyperref}
\usepackage{float}
\usepackage{booktabs}
\usepackage{array}
\usepackage{longtable}
\usepackage{amsmath}
\usepackage{enumitem}
\usepackage{xcolor}
\usepackage{titlesec}
\usepackage{caption}
\usepackage{subcaption}
\usepackage{url}
\hypersetup{
    colorlinks=true,
    linkcolor=blue,
    citecolor=blue,
    urlcolor=blue
}
\title{Galaxy Tracer: A Topology-First 3D Interface for Interactive PCAP Exploration}
\author{
Ryan Younger \\
Department of Computer Science \\
Olivet Nazarene University \\
\texttt{ryounger@olivet.edu}
}
\date{2026}
\begin{document}
\maketitle
\begin{abstract}
Packet analysis tools conventionally present capture data through tabular packet lists, constraining the analyst to a sequential view that obscures the relational structure of network communication. This paper presents Galaxy Tracer, a browser-native packet capture exploration system in which the default interface is an interactive three-dimensional network topology rather than a packet list. Hosts appear as spatially positioned nodes, conversations as edges, and protocol groupings as visually distinct clusters. A synchronized packet list remains available as a secondary view, sharing filter state with the topology so that structural and tabular inspection function as one continuous workflow. The system parses PCAP and PCAPNG formats, dissects over 90 protocols, and renders the topology through Three.js. The paper argues that the third spatial dimension is not merely aesthetic but analytically meaningful: it reveals density, clustering, host centrality, and communication scale that are difficult to perceive in list-only tools. System capabilities and performance characteristics are summarized in Table~\ref{tab:summary}.
\end{abstract}
\section{Introduction}
Packet capture analysis remains a foundational task in networking, cybersecurity, digital forensics, systems administration, and network education. Packet captures preserve exact communication events and therefore support deep protocol inspection, troubleshooting, intrusion analysis, and teaching. In practice, however, packet captures are most commonly explored through tabular packet lists in which each row corresponds to a packet and additional panes expose protocol fields, flags, and payload structure.
This design has proven extraordinarily effective for expert analysis. Yet it also imposes perceptual limitations. Large captures remain visually compressed into a narrow list pane. The analyst can see only a small subset of rows at once, and the global structure of network communication must be mentally reconstructed from repeated source and destination addresses. A capture containing dozens of packets, therefore, often appears little different from one containing tens of thousands apart from the length of the vertical scrollbar.
Galaxy Tracer was developed in response to this limitation. It proposes a topology-first approach to packet capture exploration in which the default mode of interaction is not a packet list but a three-dimensional spatial topology of communication. Hosts, flows, and protocol groupings are rendered as an interactive graph-like structure that reveals density, clustering, and communication patterns immediately. A synchronized packet list remains available for detailed inspection, but it is no longer the user's point of entry.
This paper makes four contributions. First, it introduces the concept of a topology-first PCAP exploration environment in which the 3D topology view is primary and the packet list is secondary. Second, it describes the design and implementation of Galaxy Tracer as a browser-native system that integrates packet parsing, shared filtering state, and interactive 3D rendering. Third, it argues that the third spatial dimension is analytically meaningful rather than merely decorative because it reveals density, clustering, and structural scale that are difficult to perceive in linear packet lists. Fourth, it provides a performance characterisation of the system including render speed, filter responsiveness, protocol coverage, and large-capture behavior (Table~\ref{tab:summary}).
The motivation for Galaxy Tracer is especially strong in educational and presentation contexts. Instructors frequently demonstrate packet captures using traditional analyzers even though the list-based interface can appear visually dense, procedural, and dull to beginners. A topology-first representation offers a more immediate and engaging overview of communication structure before students descend into packet-level detail. Similarly, in professional settings, a structural visualization may communicate network behavior to stakeholders more effectively than rows of source and destination fields alone. The current implementation also includes a built-in synthetic startup demo so that users can begin interacting with the interface immediately before loading an external capture.
\section{Problem Statement}
Traditional packet analysis interfaces are optimized for precise inspection. They excel at answering questions such as which flag was set, what field value appeared in a specific packet, or which byte sequence was present in a payload. These are indispensable forensic and engineering tasks. However, the same interfaces are less effective when the user seeks to understand the broader shape of a capture.
Several recurring limitations follow.
\begin{enumerate}[label=\arabic*.]
    \item \textbf{Structural opacity.} Communication relationships between hosts are not directly visible. The analyst must infer hubs, fan-out patterns, and grouped behavior by scanning repeated addresses across rows.
    \item \textbf{Compressed scale.} The visual difference between a modest capture and a very large capture is often surprisingly small in a single-pane list view. Scale is represented indirectly rather than perceptually.
    \item \textbf{Cognitive reconstruction overhead.} The user must mentally reconstruct a graph from a sequence of rows. This is especially burdensome for beginners and intermediate learners.
    \item \textbf{Low demonstration legibility.} Dense packet lists are effective for experts but can be less effective for projection, teaching, and non-specialist communication.
    \item \textbf{Reduced exploratory appeal.} Conventional interfaces privilege procedural inspection over curiosity-driven exploration.
\end{enumerate}
These limitations do not imply that list-based analyzers are flawed. Rather, they show that packet capture data supports multiple useful representations and that the dominant tabular representation is only one of them. Network traffic is inherently relational. It consists of communicating endpoints, protocol groupings, repeated flows, asymmetries, hubs, and clusters. These properties are naturally graph-like. A topology-oriented visualization can therefore expose meaningful features of the data that are latent but difficult to perceive in a list.
Galaxy Tracer addresses this by making topology the primary mode of analysis. The user begins with communication structure and descends into rows only when detailed inspection is needed.
\section{Related Work}
Packet analysis tools have historically centered on protocol dissection, packet filtering, and field-level inspection. The most influential example is Wireshark, which remains the dominant open network protocol analyzer and provides a mature workflow for packet-level inspection, filtering, and dissection \cite{wireshark}. Wireshark establishes the practical baseline for packet analysis in both education and professional practice. It also includes several visualization features beyond its core packet list, including I/O graphs, flow sequence diagrams, and an endpoint map that plots IP addresses geographically using MaxMind GeoIP databases. These features are valuable but exist as secondary windows invoked from a list-first workflow. They do not change the fundamental interaction model: the packet list remains the default and primary surface, and the visualizations serve as auxiliary views. Galaxy Tracer's contribution is to invert this relationship, making topology the default interface rather than an optional supplement.
At a lower level, packet capture itself is commonly supported by libpcap and related capture libraries, which form the basis for many packet-oriented tools and workflows \cite{pcap}. These libraries enable packet acquisition and decoding pipelines but do not determine the higher-level interaction model by which captures are explored.
Several tools have explored network-specific visualization. EtherApe renders network activity as a real-time graph of host communication, providing a topology view driven by live traffic or offline captures \cite{etherape}. NetworkMiner provides a host-centric view of PCAP data with extracted artifacts and communication summaries \cite{networkminer}. RUMINT offers multiple visual representations of packet captures including parallel coordinate plots and scatter displays \cite{rumint}. These tools demonstrate that topology and structural visualization of network traffic is a recognized and productive direction. However, they typically operate as standalone visualization utilities or forensic tools rather than as integrated environments in which topology and packet-level inspection share filtering state and can be switched between immediately. Galaxy Tracer differs in combining direct PCAP ingestion, a topology-first default interface, a synchronized secondary packet list view, and shared filtering in one browser-native system.
More broadly, network visualization research has explored graph layouts, flow diagrams, topology views, and visual analytics for security and infrastructure data.
Ware and Bobrow demonstrated that dynamic motion cues can support rapid interactive queries on node-link diagrams, highlighting the value of perceptual feedback in graph-based interfaces \cite{warebobrow}. Herman, Melancon, and Marshall surveyed graph visualization and emphasized that network structure often benefits from representations that expose clustering, path relationships, and topology rather than sequential detail alone \cite{herman}. Similarly, Munzner's work on graph visualization and visual abstraction established that relational datasets often require different visual encodings than tabular data \cite{munzner}.
In security visualization, Goodall, Lutters, and Komlodi showed that visual analytics can support understanding of network events by exposing higher-level patterns that would otherwise be difficult to infer from raw records alone \cite{goodall}. Conti's work on security data visualization likewise argued that human analysts benefit from representations that reveal pattern, anomaly, and relationship rather than only textual detail \cite{conti}. These lines of work strongly support the premise that packet-derived network behavior can profit from visualization beyond the conventional packet list.
Galaxy Tracer's interaction model also draws on established principles from human-computer interaction research. Shneiderman's Visual Information Seeking Mantra---overview first, zoom and filter, then details on demand \cite{shneiderman}---directly informs the system's design: the topology provides the overview, protocol filtering narrows the view, and the packet list supplies detail. Yi, Kang, Stasko, and Jacko identified seven categories of interaction for information visualization, including filtering, selecting, and connecting, all of which Galaxy Tracer supports through its clickable legend, node expansion, and synchronized views \cite{yi}. Elmqvist and Fekete's work on hierarchical aggregation in information visualization \cite{elmqvist} is relevant to Galaxy Tracer's design decision to cap the topology at the 80 most active hosts---a form of aggregation that preserves readability while summarizing the full dataset. Nielsen's usability heuristic of recognition rather than recall \cite{nielsen} further supports the topology-first approach: the spatial layout allows users to recognize communication patterns visually rather than reconstructing them mentally from sequential rows.
Three-dimensional rendering frameworks such as Three.js make browser-native spatial interfaces feasible and have lowered the barrier to interactive visualization on commodity hardware \cite{threejs}. However, the existence of 3D rendering frameworks does not by itself solve the packet analysis problem. The key issue is not whether packets can be drawn in 3D, but whether a topology-first representation can function as a practical and synchronized exploratory interface for real PCAP data.
\section{Design Goals}
The design of Galaxy Tracer was guided by five goals.
\subsection{Topology First}
The first goal was to invert the conventional packet analysis workflow. Instead of treating topology as an optional visualization mode, Galaxy Tracer makes topology the default and primary analytic surface.
\subsection{Shared Analytic State}
The second goal was to ensure that topology and packet list views behave as two renderings of the same underlying state rather than as separate tools. Filtering and selection should persist immediately across both views.
\subsection{Immediate Scale Perception}
The third goal was to reveal the scale and density of a capture perceptually. Large captures should feel large. Dense captures should look dense. Sparse captures should look sparse.
\subsection{Exploratory Interaction}
The fourth goal was to support curiosity-driven exploration. If a user filters by protocol, isolates traffic, or pivots views, the result should be immediate and visually meaningful.
\subsection{Browser-Native Delivery}
The fifth goal was practical portability. The system should run in a browser without requiring a traditional desktop installation, thereby supporting classrooms, touch devices, rapid demonstration, and lightweight distribution.
\section{System Architecture}
Galaxy Tracer consists of four principal layers: capture ingestion, parsing and normalization, graph construction, and view rendering.
\subsection{Capture Ingestion}
The system accepts packet capture input through file selection and drag-and-drop. Format detection uses capture magic bytes rather than filename extension, supporting both legacy PCAP and PCAPNG formats. Unsupported formats are rejected with diagnostic feedback including the detected magic bytes. The parser currently enforces a safeguard of 100{,}000 packets.
\subsection{Parsing and Normalization}
Packet data is translated into a shared internal model suitable for both topology and tabular rendering. Rather than maintaining independent pipelines for each view, Galaxy Tracer constructs a single in-memory representation that can be queried by either renderer. The packet list uses virtual scrolling so that only visible rows and a small overscan region are mounted in the DOM at any one time. In the PCAPNG parsing path, control is periodically yielded back to the browser via \texttt{requestAnimationFrame} so that interface responsiveness is preserved during longer parses.
This shared-model design is important because it permits immediate switching between views. The packet list and topology are not separate analyses performed sequentially. They are two projections of the same parsed dataset.
\subsection{Graph Construction}
The topology renderer derives nodes and edges from the shared internal model. Communication entities are represented as spheres in 3D space, while relationships between them are represented as connecting edges whose visual weight reflects traffic volume. In the current implementation, topology edges are constructed from communication pairs and the rendered topology is intentionally limited to the most active hosts for readability. Hosts are ranked by packet count and only the top 80 are visualized. The purpose of the graph is not to replace packet-level detail but to reveal communication structure, including concentration, clustering, and centrality.
\subsection{View Rendering}
Galaxy Tracer currently exposes two synchronized views.
\begin{enumerate}[label=\arabic*.]
    \item \textbf{3D topology view.} The default and primary interface. This is the main analytical surface and the central novelty of the system.
    \item \textbf{Packet list view.} A secondary interface for traditional packet inspection.
\end{enumerate}
Because both views operate on the same internal state, changing a filter in one immediately affects the other. The user is therefore not switching tools but changing analytic perspective.
\begin{figure}[H]
    \centering
    \includegraphics[width=0.95\linewidth]{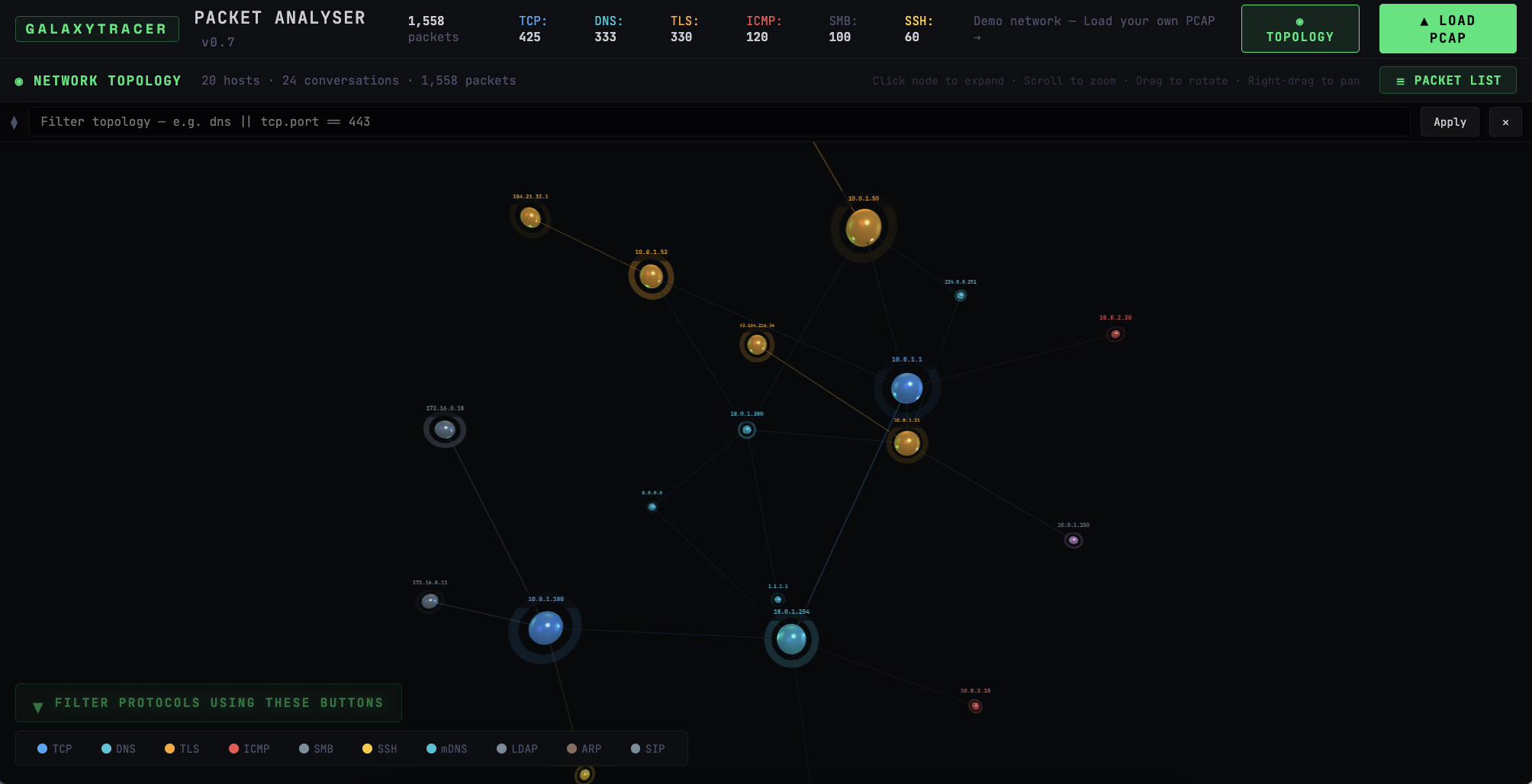}
    \caption{Galaxy Tracer in its default topology-first mode showing the built-in demo network: 20 hosts, 24 conversations, and 10 distinct protocols.}
    \label{fig:topology_overview}
\end{figure}
\section{Interaction Model}
A central contribution of Galaxy Tracer lies in its interaction model rather than in graphics alone.
\subsection{Default Topology View}
Upon opening the application, the user lands directly in the 3D topology environment. This immediately establishes the tool's identity as a topology-first analyzer rather than a packet list with an optional visual mode. If no external capture has yet been loaded, the application presents a built-in synthetic demo so that the user can explore the interaction model immediately.
\subsection{Packet List as Secondary Lens}
The packet list remains available for detailed inspection, but it is no longer the user's starting point. The system first presents the structure of communication and only secondarily the sequential detail of packets.
\subsection{Text Filtering}
A filter bar allows the user to restrict the visible subset of traffic using Wireshark-style display filter syntax. This filter state is shared between views. Entering a filter in the topology view changes the displayed structure, and switching to the packet list then shows the matching packet subset immediately.
\subsection{Clickable Protocol Legend}
The protocol legend is interactive and dynamic, showing only protocols present in the current capture, sorted by frequency. Selecting a protocol such as DNS filters the topology to that subset. This turns what would otherwise be a passive legend into an active control surface and shortens the loop between curiosity and exploration.
\subsection{Touch Support and Micro-Interaction}
Touch and mobile support were added so that the topology can be manipulated on tablets and touch-enabled devices. The current implementation supports single-touch orbit, two-finger pinch zoom, two-finger pan, and tap-to-select interaction on topology elements. Animation feedback on interaction, including eased node expansion and glow-ring breathing, helps the tool feel responsive and instrument-like rather than static.
\subsection{Immediate View Switching}
Because the system maintains a shared in-memory representation, switching between topology and packet list is immediate (see Table~\ref{tab:summary}). This preserves the user's mental model of the data and allows structural and tabular inspection to function as one continuous workflow.
\begin{figure}[H]
    \centering
    \includegraphics[width=0.95\linewidth]{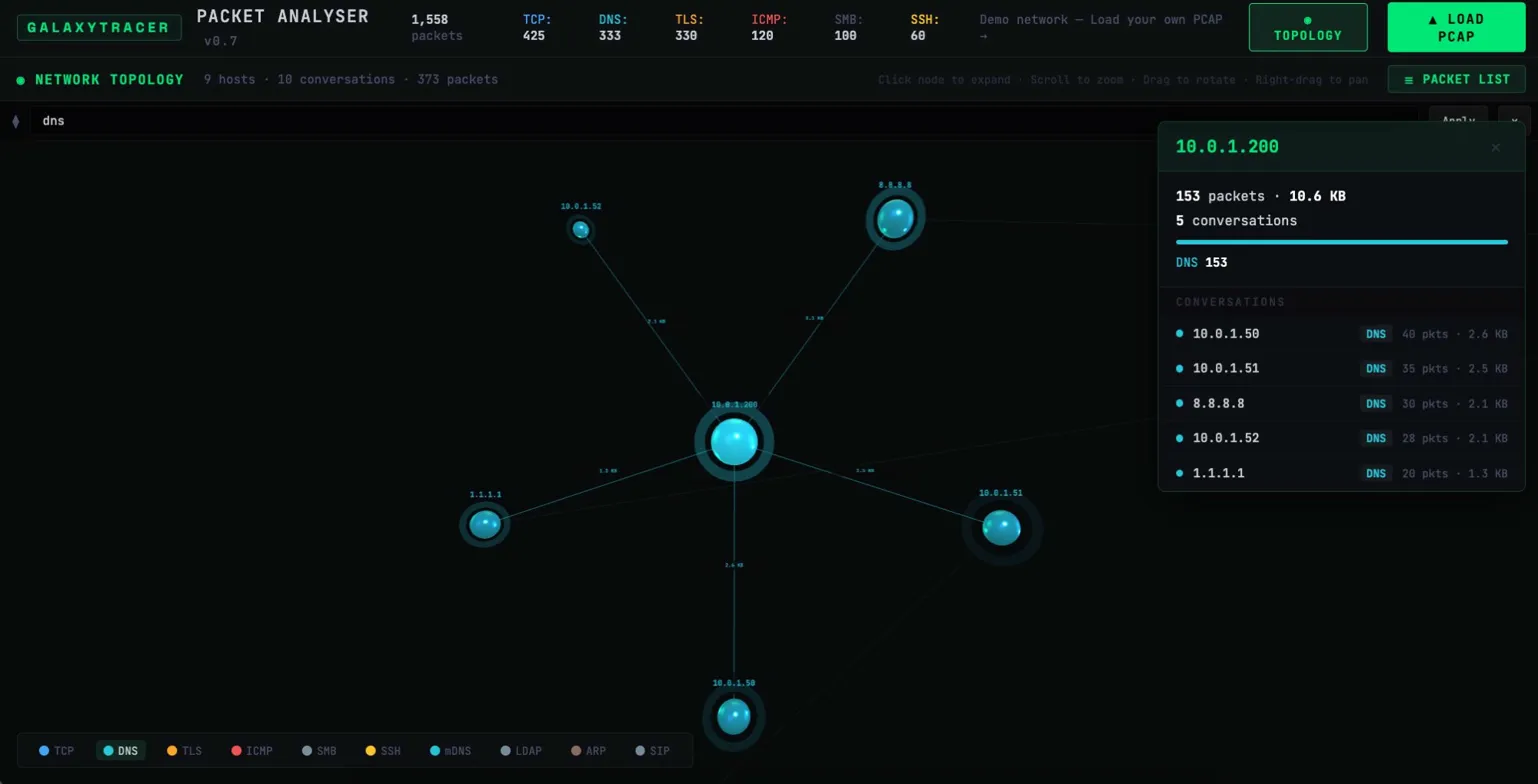}
    \caption{DNS filter applied with node 10.0.1.200 expanded, revealing its role as internal
DNS server: three workstation clients (10.0.1.50, 10.0.1.51, 10.0.1.52) and two upstream
resolvers (8.8.8.8 and 1.1.1.1) are visible simultaneously. The conversation panel
confirms 153 DNS packets across 5 conversations.}
    \label{fig:dns_filter}
\end{figure}
\section{Why the Third Dimension Matters}
Three-dimensional visualization is often criticized because in many contexts it adds novelty without informational value. In charting and dashboard environments this criticism is frequently justified. Decorative depth can reduce clarity rather than improve it.
Galaxy Tracer uses the third dimension differently. The system is not adding depth to a flat chart for visual flourish. It is spatializing a relational dataset. Network traffic is inherently graph-like, consisting of communicating entities, repeated flows, clustered behavior, and dense or sparse regions of interaction.
The 3D topology offers several analytical advantages.
\subsection{Scale Becomes Visible}
In a packet list, a capture containing dozens of packets and one containing tens of thousands may appear visually similar apart from the length of the vertical scrollbar. In the topology view, larger and denser captures produce visibly larger or denser structures within the rendered active-host subset. Scale therefore becomes perceptual rather than merely quantitative, even though the current implementation intentionally summarizes the topology rather than rendering every endpoint without bound.
\subsection{Density Becomes Spatial}
Traffic density becomes visual mass. Heavily connected or highly trafficked regions of a capture appear denser in space, while sparse captures remain sparse.
\subsection{Clustering and Centrality Become Immediate}
Communication hubs, repeated service relationships, grouped protocol activity, and fan-out patterns become easier to perceive directly.
\subsection{Protocol Isolation Reveals Structure}
When a protocol such as DNS is selected, the graph transforms to reveal the structural subset associated with that protocol. The user can see what that protocol is doing in the network rather than merely listing matching rows.
As a concrete example, when filtering to DNS traffic in the built-in demo network and
expanding the central host, the topology immediately reveals that 10.0.1.200 acts as the
internal DNS server, maintaining conversations with three workstation clients
(10.0.1.50, 10.0.1.51, and 10.0.1.52) and forwarding queries upstream to two external
resolvers (8.8.8.8 and 1.1.1.1). All five relationships are visible simultaneously in the
topology. Identifying the same structure in a conventional packet list requires the analyst
to scroll through hundreds of rows, mentally tracking repeated source-destination pairs
across non-contiguous entries.
\subsection{Demonstration Quality Improves}
In teaching and presentation contexts, the topology view provides a far more immediate and legible high-level representation of traffic structure than a dense packet list.
For these reasons, the third dimension in Galaxy Tracer should be understood as an analytical projection rather than an aesthetic layer.
\section{Implementation Status}
In the current implementation, Galaxy Tracer includes the following features:
\begin{itemize}
    \item topology-first default view,
    \item synchronized secondary packet list view with virtual scrolling,
    \item text-based traffic filtering with Wireshark-style display filter syntax,
    \item dynamic clickable protocol legend,
    \item touch support including orbit, pinch zoom, and pan,
    \item animation feedback including eased expansion and glow-ring breathing,
    \item a built-in synthetic startup demo for immediate engagement,
    \item drag-and-drop and file-input capture loading,
    \item browser-native Three.js rendering with PBR materials and environment mapping,
    \item PCAP and PCAPNG parsing with format detection by magic bytes,
    \item protocol dissection covering over 90 protocols,
    \item node expansion with conversation detail panel and edge traffic labels,
    \item right-click-drag camera panning,
    \item a parser safeguard of 100{,}000 packets,
    \item a topology view limited to the top 80 hosts by packet count for readability.
\end{itemize}
The implementation is sufficient for live classroom demonstration and exploratory packet analysis. The author continues to refine the interface incrementally through regular use and teaching practice.
\section{Evaluation}
Galaxy Tracer is evaluated as a systems tool. The most important signals are performance, responsiveness, protocol coverage, and behavior at implementation limits. Table~\ref{tab:summary} consolidates capability and performance data for the current implementation. Timing measurements were collected on the benchmark capture, The Ultimate PCAP v20251206 (48{,}640 packets, 14.1~MB, containing traffic from over 90 distinct protocols). Each timing metric was measured as the mean of three consecutive runs using browser developer tools on the test platform specified in the table.
\begin{table}[H]
\centering
\caption{Galaxy Tracer: system capability and performance summary.}
\begin{tabular}{ll}
\toprule
\textbf{Metric} & \textbf{Value} \\
\midrule
Supported capture formats & PCAP, PCAPNG (detected by magic bytes) \\
Supported protocols & 90+ \\
Parser safeguard & 100{,}000 packets \\
Topology node cap & 80 hosts \\
Benchmark capture & 48{,}640 packets (14.1~MB) \\
Time to interactive state (benchmark) & $\sim$0.4 s \\
Protocol filter latency (benchmark) & $\sim$0.2 s \\
View switch latency (benchmark) & $\sim$0.2 s \\
Test platform & MacBook Air M4 (2025), 16 GB RAM, macOS 15.7.3 \\
Browser & Google Chrome 145.0.7632.117 (Official Build) (arm64) \\
\bottomrule
\end{tabular}
\label{tab:summary}
\end{table}
\section{Educational and Demonstration Value}
Galaxy Tracer was developed in part as a supplementary teaching instrument for packet analysis. This educational context strongly influenced the design. In classroom settings, traditional analyzers are often effective for labs but less effective for first-contact demonstrations. Students are shown dense packet lists before they have formed a mental model of the communication structure. This can make packet analysis appear procedural and visually dull.
A topology-first demonstration changes that dynamic. An instructor can begin by showing the structural shape of a capture, then use the packet list only when detail is required. In this sense, Galaxy Tracer complements existing analyzers rather than replacing them. It offers a structural overview first and a packet microscope second.
The system also has value as a presentation tool. Instructors, network engineers, and analysts may find that a topology view communicates communication structure more effectively to audiences than rows of packet metadata alone.
\section{Positioning and Scope}
Galaxy Tracer is not presented here as a universal replacement for mature packet analyzers. Expert analysts will still require scripting, deep dissectors, payload inspection, and advanced forensic workflows.
Instead, Galaxy Tracer is positioned as a topology-first companion or alternative lens for beginner to intermediate users, for educational environments, and for exploratory structural analysis. It is especially relevant where the immediate perception of scale and communication structure matters.
This positioning is a strength rather than a weakness. Many successful technical tools thrive by serving a specific use case exceptionally well rather than attempting to replace every incumbent system.
\section{Future Work}
Several extensions are possible. Larger-scale testing and additional benchmarks should be conducted. Protocol coverage can continue to expand. Additional visual analytics such as anomaly highlighting, host grouping, and conversation clustering may be introduced. Synthetic packet generation or packet crafting environments could eventually support controlled topology creation for teaching and testing. IP geolocation mapping, in which the same node data is projected onto a 3D globe rather than a force-directed graph, represents another natural extension of the spatial approach.
\section{Conclusion}
Galaxy Tracer introduces a topology-first approach to packet capture exploration. The 3D topology view is the primary interface; the packet list is secondary. This inversion allows users to perceive communication structure, scale, and protocol segmentation in ways that list-only tools do not make immediately visible. The third dimension is not decorative. It is an analytical projection that reveals density, clustering, and centrality in relational network data. System capabilities and measured performance are summarized in Table~\ref{tab:summary}. Galaxy Tracer demonstrates that packet capture analysis need not begin with rows. It can begin with structure.

\par\bigskip
\noindent\textbf{Availability.} Galaxy Tracer is hosted at \url{https://ryanyounger.ai/galaxytracer}. Source code is not currently publicly released.
\bibliographystyle{plain}

\end{document}